\documentclass[aps,preprint,nofootinbib,eqsecnum]{revtex4}%
\usepackage{amsmath}
\usepackage{amsfonts}
\usepackage{amssymb}
\usepackage{graphicx}
\usepackage{amscd}%
\providecommand{\U}[1]{\protect\rule{.1in}{.1in}}

\allowdisplaybreaks
\begin{document}
\leftline {USC-10/HEP-B3 \hfill CERN-PH-TH/2010-171}{}{\vskip-1cm}

{\vskip2cm}

\begin{center}
{\Large \textbf{Constraints on Interacting Scalars in 2T Field Theory }}

{\Large \textbf{and No Scale Models in 1T Field Theory}}\footnote{This work
was partially supported by the US Department of Energy under grant number
DE-FG03-84ER40168.}{\Large \textbf{\ }}

{\vskip0.4cm}

\textbf{Itzhak Bars}

{\vskip0.4cm}

\textsl{Department of Physics and Astronomy}

\textsl{University of Southern California,\ Los Angeles, CA 90089-0484 USA}

and

\textsl{Theory Division, Physics Department, CERN, 1211 Geneva 23,
Switzerland}

{\vskip1.0cm} \textbf{Abstract}
\end{center}

In this paper I determine the general form of the physical and mathematical
restrictions that arise on the interactions of gravity and scalar fields in
the 2T field theory setting, in d+2 dimensions, as well as in the emerging
shadows in d dimensions. These constraints on scalar fields follow from an
underlying Sp$\left(  2,R\right)  $ gauge symmetry in phase space. Determining
these general constraints provides a basis for the construction of 2T
supergravity, as well as physical applications in 1T-field theory, that are
discussed briefly here, and more detail elsewhere. In particular, no scale
models that lead to a vanishing cosmological constant at the classical level
emerge naturally in this setting.

\section{Prologue: The role of the Sp(2,R) gauge symmetry}

The progress of two-time physics (2T-physics), from classical mechanics to the
standard model, supersymmetric 2T field theory, 2T general relativity and
cosmology, is summarized in \cite{phaseSpace}. More recently, 2T field theory
computational techniques for 2T-physics that are directly in 4+2 dimensions
have emerged \cite{rivelles}\cite{weinberg} and mathematical fields that
adopted the notions of 2T-physics have advanced \cite{waldron}.

2T-physics is the fundamental solution to the question of how to make a theory
with two timelike dimensions causal, ghost-free and unitary directly in
d-space and 2-time dimensions. Field theories based on the principles of
2T-physics in d+2 dimensions, including the standard model and general
relativity, automatically have the right mix of gauge symmetries and
constraints to make them compatible with their conventional formulation in
field theory in 1T-physics in 3+1 dimensions. Conventional classical or
quantum mechanics as well as 1T field theory emerge as \textquotedblleft
shadows\textquotedblright\ \cite{phaseSpace} of 2T-physics systems in one less
time and one less space dimensions. Beyond this compatibility with 1T-physics,
2T-physics predicts systematically additional physical information, in the
form of hidden symmetries and dualities among the shadows, that is
systematically missed information in the conventional 1T formulation. One of
these shadows, called the \textquotedblleft conformal shadow\textquotedblright%
\ is the one most familiar to particle physicists. In this shadow conformal
symmetry SO(4,2) is a 4-dimensional non-linear realization of the linear
Lorentz symmetry in 4+2 dimensions. The new aspects of 2T-physics follow from
demanding a fundamental Sp(2,R) gauge symmetry in phase space, which in turn
requires a primary ambient spacetime, with no more and no less than two
timelike dimensions, in which conventional spacetime is embedded.

The role of the fundamental Sp$\left(  2,R\right)  $ gauge symmetry in
2T-physics is similar to the worldsheet gauge symmetry in string theory. In
the case of Sp$\left(  2,R\right)  $ there are three gauge generators,
$Q_{ij}(X,P),$ $i,j=1,2,$ arranged into a $2\times2$ symmetric matrix$,Q_{11}%
,Q_{22},Q_{12}=Q_{21},$ constructed from phase space degrees of freedom on the
worldline $X^{M}\left(  \tau\right)  ,P_{M}\left(  \tau\right)  ,$ with a
target spacetime in $d+2$ dimensions labelled by $M.$ Comparing to string
theory, the $Q_{ij}(X,P)$ are analogous to the Virasoro operators constructed
from string coordinates $x^{\mu}\left(  \tau,\sigma\right)  $ and its
derivatives (or canonical oscillators, i.e. phase space). In both cases the
particle or string may interact with an infinite set of background fields
(electromagnetism, gravity, etc.) that appear in the expressions of the
Sp$\left(  2,R\right)  $ or Virasoro generators. All such $Q_{ij}\left(
X,P\right)  $ in 2T-physics have been classified up to canonical
transformations\footnote{The simplest example of the $Q_{ij}$ is
$Q_{11}=X\cdot X,$ $Q_{12}=X\cdot P,$ $Q_{22}=P\cdot P,$ where the dot product
involves the flat background metric $\eta_{MN}$ in $d+2$ dimensions. If
$\eta_{MN}$ had been a Euclidean metric or a 1T Minkowski metric, the three
equations $Q_{ij}=0$ yield only trivial solutions. Non-trivial as well as
ghost free physics occurs only when two timelike dimensions are admitted. The
solutions are called shadows when a gauge is chosen to embed 3+1 phase space
in 4+2 phase space (see Figs.1.2 in \cite{phaseSpace}). Because many different
embeddings exist such that \textquotedblleft time\textquotedblright\ and
\textquotedblleft Hamiltonian\textquotedblright\ are identified differently as
part of 4+2 phase space, the 1T-physics is different in different 3+1 shadows,
as seen by observers restricted to those shadows. Nevertheless many relations
do exist among the shadows. This is the new information predicted by
2T-physics that is systematically missed in 1T-physics. For each choice of
$Q_{ij}(X,P),$ including background fields \cite{2tbacgrounds}, there is a
corresponding set of shadows analogous to Fig.1 in \cite{phaseSpace}. Hence
the systematically missed hidden information in 1T physics is vast.
\label{simple}} \cite{2tbacgrounds}.

Demanding Sp$\left(  2,R\right)  $ gauge symmetry on the worldline is similar
to demanding local conformal symmetry on the worldsheet. To preserve the
Sp$\left(  2,R\right)  $ symmetry, the background fields that appear in
$Q_{ij}(X,P)$ must satisfy certain equations that follow from requiring
closure of the $Q_{ij}(X,P)$ as the Sp$\left(  2,R\right)  $ Lie algebra under
Poisson brackets. The resulting field equations are analogous to the equations
for background fields in string theory that follow from demanding exact
worldsheet conformal symmetry. However, in the case of Sp$\left(  2,R\right)
$ these are \textquotedblleft kinematic equations\textquotedblright, that
guarantee the absence of ghosts, without determining the dynamics of the
background fields. To determine the dynamics, consistently with the kinematics
imposed by Sp$\left(  2,R\right)  ,$ a unique action principle that has its
own gauge symmetries is constructed as illustrated in \cite{2tGravity}%
\cite{2tGravDetails}. In this paper I will discuss the extent to which this
approach permits generalizations while still imposing constraints on the
interactions of scalar fields in 2T field theory.

Since the $Q_{ij}(X,P),$ like the Virasoro operators, are gauge generators,
the gauge invariant space (at the particle level) is defined as the subspace
of phase space that satisfies $Q_{ij}(X,P)=0.$ As can be verified through the
example in footnote (\ref{simple}), there is non-trivial physics content in
the equations $Q_{ij}(X,P)=0$ provided there are two timelike dimensions in
target space $X^{M}$ \cite{phaseSpace}. Furthermore, the available gauge
symmetry is just right to eliminate the ghosts of two timelike dimensions.
Therefore unitarity at the quantum mechanics level (wavefunctions
$\leftrightarrow$ fields) is achieved only for a spacetime with two times, no
less and no more.

Unitarity is also insured for spinning particles on a $d+2$ dimensional target
space (analogs of Neveu-Schwarz degrees of freedom in string theory) or
supersymmetric systems (analogs of Green-Schwarz degrees of freedom in string
theory) by enlarging the gauge symmetry to OSp$\left(  n|2\right)  $
\cite{spin2t}\cite{2Tfields2000} or with appropriate supercoset symmetries
\cite{super2t}-\cite{spinBO} respectively. The latter lead to novel twistor
and supertwistor representations of various systems \cite{twistorLect},
including the 2T-physics version of the twistor superstring in 4+2 dimensions.

2T field theory is constructed with similar techniques to string field theory.
The basic equations for free scalar fields are the physical state conditions
that correspond to gauge invariance under Sp$(2,R),$ $Q_{ij}(X,\partial
)\Phi\left(  X\right)  =0$ (with an appropriate ordering of $X$ and
$P=-i\partial/\partial X$)$.$ This is analogous to the Virasoro constraints
applied on string fields\footnote{For example, for a scalar field in a flat
background in $d+2$ dimensions (see footnote \ref{simple}), these equations
are $Q_{11}\Phi=X^{2}\Phi=0,$ $Q_{12}\Phi=(X\cdot\partial+\partial\cdot
X)\Phi=0$ and $Q_{22}\Phi=\partial\cdot\partial\Phi=0.$ The first two
equations are called \textquotedblleft kinematic\textquotedblright\ and the
last \textquotedblleft dynamic\textquotedblright. The solution of the first
equation is $\Phi\left(  X\right)  =\delta\left(  X^{2}\right)  \phi\left(
X\right)  ,$ indicating that $\phi\left(  X\right)  \rightarrow\phi\left(
X\right)  +X^{2}\Lambda\left(  X\right)  $ is a gauge symmetry. This also
illustrates the origin of the delta function $\delta\left(  W\right)  $ that
occurs in the actions (\ref{action},\ref{actionU}) since in flat space
$W=X^{2}$ as seen in Eq.(\ref{Qgrav}). Inserting this solution in the second
equation, and using the homogeneity of the delta function, one obtains
$\delta\left(  X^{2}\right)  \left(  2X\cdot\partial+d-2\right)  \phi\left(
X\right)  =0,$ indicating that $\phi\left(  X\right)  $ must be homogeneous of
degree $\left(  2-d\right)  /2$. Combining these properties one learns that
$\phi\left(  X\right)  $ depends really on $d$ coordinates rather than $d+2$
coordinates. The various shadows of the surviving degrees of freedom in $\Phi$
correspond to how the \textit{phase space} in $d$ dimensions is embedded in
the phase space in $d+2$ dimensions. The final (shadow) form of the last
dynamical equation depends on this embedding. The conformal shadow is one of
those examples. \label{kinEqs}}. A systematic approach to include interactions
in 2T field theory is to construct the BRST operator for Sp$\left(
2,R\right)  $, use it as the kinetic operator in a BRST field theory, and then
include interactions \cite{2tbrst2006}. The inclusion of interactions modifies
the BRST operator \cite{2tbrst2006}, just as it does in string field theory.

A complementary approach to 2T field theory is a procedure analogous to the
construction of the string effective action as a field theory. It consists of
a field theory action in $d+2$ dimensions that reproduces the same equations
for the background fields that were obtained by demanding the Sp$\left(
2,R\right)  $ gauge symmetry on the worldline in the presence of backgrounds
\cite{2tGravity}\cite{2tGravDetails}. In this paper I will discuss, in this
approach, the generalizations of the gravity action in \cite{2tGravity}%
\cite{2tGravDetails} when many scalar fields are present.

One can verify that the two approaches, BRST or effective action, lead to the
same \textquotedblleft kinematical\textquotedblright\ and \textquotedblleft
dynamical\textquotedblright\ equations\ after redundant fields are integrated
out in the BRST field theory \cite{2tbrst2006}\footnote{Our BRST approach of
\cite{2tbrst2006} is directly applicable with all background fields in the
$Q_{ij}\left(  X,P\right)  $ as in \cite{2tbacgrounds}. This was adopted
recently by mathematicians in \cite{waldron} (with a limited subset of
background fields) to develop a mathematical topic called \textquotedblleft
tractors\textquotedblright. Although not discussed in \cite{waldron}, I
suggest that this mathematical topic can be further generalized by using the
BRST operators for OSp(n%
$\vert$%
2) that includes spin degrees of freedom, as in \cite{2Tfields2000} or
supersymmetric degrees of freedom as in \cite{super2t}-\cite{spinBO}. Deeper
physical and mathematical results would be obtained by deriving the kinematic
constraints from an improved action principle (as in \cite{noncommutative}) or
with a delta function (as in Eq.(\ref{action}) and \cite{2tGravity}) as
opposed to applying them as external constraints as done in \cite{waldron}.
\label{brstApproach}}. In contrast to the kinematical equations that are
algebraic or at most first order in derivatives (see e.g. $Q_{11},Q_{12}$ in
footnote \ref{kinEqs}, or see Eqs.(\ref{sp3})), the dynamical equations are
those that include Klein-Gordon or Dirac-like differential operators (such as
$Q_{22}\Phi=\partial\cdot\partial\Phi+\cdots\sim0$) whose details include also
interactions. Kinematical and dynamical equations for all spinning fields
follow from either one of the two complementary approaches as shown in
\cite{2tbacgrounds}\cite{2Tfields2000}.

These were the 2T-physics principles that led to 2T field theory for all
spinning fields. They were used to construct the standard model
\cite{2tstandardM}, supersymmetric generalizations \cite{susy2tN1}%
\cite{susy2tN2N4}, and general relativity \cite{2tGravity}\cite{2tGravDetails}
as unitary field theories in 4+2 dimensions. The action principle provides all
the necessary ingredients to perform computations directly in $d+2$ dimensions.

Recent examples of computation of n-point Green's functions in 4+2 dimensions
that obey all the \textquotedblleft kinematic equations\textquotedblright\ of
2T-physics was given by Weinberg \cite{weinberg}\footnote{In the initial
version of ref.(\cite{weinberg}) Weinberg was not aware that the constraints
he used to derive his 6-dimensional Green's functions were identical to the
\textquotedblleft kinematic equations\textquotedblright\ of 2T-physics field
theory in 4+2 dimensions. The constraints for tensors or spinors were
previously derived from an underlying principle based on Sp$\left(
2,R\right)  $ gauge symmetry \cite{2tbacgrounds}, and related extended gauge
symmetries with spin \cite{spin2t}\cite{2Tfields2000}, and these appeared in
realistic 2T-physics models in flat spacetime, including the standard model in
4+2 dimensions \cite{2tstandardM}. Furthermore, the reduction to 3+1
dimensions of the Green's functions discussed in \cite{weinberg} parallels the
corresponding reduction of the fields discussed in \cite{emergentfieldth1}%
\cite{emergentfieldth2} just for the conformal shadow. This link to conformal
symmetry, that led to the 6 dimensional equations with some educated
guesswork, goes back to Dirac \cite{Dirac}-\cite{vasiliev2}. However, until
the emergence of 2T-physics it was not realized that (i) there is a
fundamental Sp(2,R) phase space gauge principle behind these equations, or
their generalizations to all tensors or spinors, which is universal for all
physics (as reviewed in \cite{phaseSpace}), and (ii) that these 6-dimensional
equations can systematically be connected to many 1T-physics shadows in 4
dimensions, not only the conformal shadow. In fact, historically both of these
points developed without any knowledge of Dirac's work. I suggest that
Weinberg's 6-dimensional Green's functions go well beyond conformal Green's
functions in 4 dimensions. They can also be reduced to other shadows just like
in \cite{emergentfieldth1}\cite{emergentfieldth2} and \cite{rivelles}. This
technique should produce Green's functions for more complicated field theories
that correspond to the shadows described earlier in 2T-physics
\cite{phaseSpace}. This is an extension to Green's functions of a similar
remark often mentioned in my 2T-physics papers as being a computational
technique that may produce non-perturbative information in 1T field theory.
\label{weinbergNote}}. Independent computations in \cite{rivelles} provide
some examples of shadows in the context of Green functions, which are
consistent with the shadows discussed before in classical or quantum mechanics
(see e.g. \cite{phaseSpace}) as well as directly in field theory
\cite{emergentfieldth1}\cite{emergentfieldth2}. I expect that such notions and
techniques that are inherently in $d+2$ dimensions are bound to open new
avenues of computation as well as provide a deeper view of space and time.

\section{2T-gravity action and physical considerations\label{GravAction}}

The 2T-physics description of a particle moving in a gravitational background
at the classical mechanics level is given by the worldline action
\cite{phaseSpace}
\begin{equation}
S=\int d\tau\left(  \dot{X}^{M}P_{M}-\frac{1}{2}A^{ij}Q_{ij}\left(
X,P\right)  \right)  ,
\end{equation}
with the following realization of the Sp$\left(  2,R\right)  $ generators
\cite{2tbacgrounds}\cite{2tGravity}
\begin{equation}
Q_{11}=W\left(  X\right)  ,~Q_{12}=V^{M}\left(  X\right)  P_{M},\;Q_{22}%
=G^{MN}\left(  X\right)  P_{M}P_{N}. \label{Qgrav}%
\end{equation}
For comparison see the flat space version in footnote (\ref{simple}) and the
more general cases in \cite{2tbacgrounds}. Closure of the Sp$\left(
2,R\right)  $ algebra under Poisson brackets is obtained, e.g. $\left\{
Q_{11},Q_{12}\right\}  =2Q_{11},$ etc., when the background fields $W\left(
X\right)  ,V^{M}\left(  X\right)  ,G^{MN}\left(  X\right)  $ satisfy the
following \textquotedblleft kinematical\textquotedblright\ equations, which
amount to homothety conditions on the geometry described by the metric
$G^{MN}$ \cite{2tbacgrounds}\cite{2tGravity}
\begin{equation}
V^{M}=\frac{1}{2}G^{MN}\partial_{N}W,\;\;V^{M}\partial_{M}%
W=2W,\;\;\pounds _{V}G^{MN}=-2G^{MN}. \label{sp3}%
\end{equation}
Here $\pounds _{V}G^{MN}$ is the Lie derivative of the metric,
\begin{equation}
\pounds _{V}G^{MN}\equiv-\nabla^{M}V^{N}-\nabla^{N}V^{M}=V^{K}\partial
_{K}G^{MN}-\partial_{K}V^{M}G^{KN}-\partial_{K}V^{N}G^{MK}.
\end{equation}
These coupled non-linear geometrical equations imply uniquely that $G_{MN}$
can be written as \cite{2tGravity}
\begin{equation}
G_{MN}=\nabla_{M}V_{N}=\frac{1}{2}\left(  \partial_{M}\partial_{N}%
W+\Gamma_{MN}^{P}\partial_{P}W\right)  . \label{sp5}%
\end{equation}
The kinematic equations (\ref{sp3}-\ref{sp5}) that follow from Sp$\left(
2,R\right)  $ are the crucial constraints that remove ghosts from the metric
degrees of freedom $G^{MN}\left(  X\right)  $ in a 2T spacetime. They can be
solved exactly \cite{2tGravity}\cite{2tGravDetails}, showing that what remains
undetermined in $G_{MN}\left(  X\right)  $ in $d+2$ dimensions are only the
degrees of freedom and gauge symmetries of a conventional shadow metric
$g_{\mu\nu}\left(  x\right)  $ in $d$ dimensions, plus gauge degrees of
freedom and prolongations of the shadow that are immaterial to the 1T physical
phenomena in the shadow \cite{2tGravity}\cite{2tGravDetails}. The dynamics of
the surviving shadow $g_{\mu\nu}\left(  x\right)  $ in $d$ dimensions (plus
matter degrees of freedom) is determined by the dynamical equations in $d+2$
dimensions that follow from a 2T field theory action principle as discussed below.

As outlined in the previous section, the homothety conditions (\ref{sp3}%
-\ref{sp5}) on the background fields are analogous to those obtained in string
theory by demanding conformal symmetry on the worldsheet. So, with a similar
approach to string theory, one can construct an effective action principle
that yields the same set of equations (\ref{sp3}-\ref{sp5}) by using the
variational principle. I proposed such an action in \cite{2tGravity} by
including one additional field $\Omega\left(  X\right)  $ that I called the
dilaton\footnote{The field $\Omega$ here is rescaled compared to
ref.\cite{2tGravity}, namely $\Omega_{old}=\sqrt{a}\Omega.$},
\begin{equation}
S=\gamma\int d^{d+2}X~\sqrt{G}\left[
\begin{array}
[c]{c}%
\delta\left(  W\right)  \left\{  a_{d}\Omega^{2}R\left(  G\right)  +\frac
{1}{2}\partial\Omega\cdot\partial\Omega-V\left(  \Omega\right)  \right\} \\
+\delta^{\prime}\left(  W\right)  \left\{  a_{d}\Omega^{2}\left(  4-\nabla
^{2}W\right)  +a_{d}\partial W\cdot\partial\Omega^{2}\right\}
\end{array}
\right]  .~ \label{action}%
\end{equation}
Here the constant $a_{d}=\frac{\left(  d-2\right)  }{8\left(  d-1\right)  }$
for $d+2$ dimensions, as well as the form of the action, are uniquely
determined by the requirement that the variation of this action reproduces the
kinematic equations (\ref{sp3}-\ref{sp5}) \cite{2tGravity}. The potential
energy is also fixed up to the overall $\lambda$ which is a dimensionless
coupling constant $V\left(  \Omega\right)  =\lambda\frac{d-2}{2d}\Omega
^{\frac{2d}{d-2}}$.

Note the unusual but crucial delta function $\delta\left(  W\left(  X\right)
\right)  $ and its derivative. In this action $W\left(  X\right)  $ is treated
as a field that is varied just as the other fields $\Omega\left(  X\right)
,G^{MN}\left(  X\right)  $ are varied. The variation of the action with
respect to each field takes the following form
\begin{equation}
\delta S=\int d^{d+2}X~\sqrt{G}\left\{
\begin{array}
[c]{c}%
\delta G^{MN}[\delta\left(  W\right)  A_{MN}^{G}+\delta^{\prime}\left(
W\right)  B_{MN}^{G}+\delta^{\prime\prime}\left(  W\right)  C_{MN}^{G}]\\
+\delta\Omega\lbrack\delta\left(  W\right)  A^{\Omega}+\delta^{\prime}\left(
W\right)  B^{\Omega}+\delta^{\prime\prime}\left(  W\right)  C^{\Omega}]\\
+\delta W[\delta^{\prime}\left(  W\right)  A^{W}+\delta^{\prime\prime}\left(
W\right)  B^{W}+\delta^{\prime\prime\prime}\left(  W\right)  C^{W}]
\end{array}
\right\}  , \label{vary}%
\end{equation}
where the coefficients of the delta functions $\left(  A_{MN}^{G},B_{MN}%
^{G},C_{MN}^{G}\right)  ,$ $\left(  A^{\Omega},B^{\Omega},C^{\Omega}\right)  $
and $\left(  A^{W},B^{W},C^{W}\right)  $ are explicitly given in
\cite{2tGravity}. All of these coefficients must vanish in order to minimize
the action since $\delta G^{MN},\delta\Omega,\delta W$ are arbitrary while the
delta function and its derivatives are linearly independent distributions. The
equations derived from the coefficients of $\delta^{\prime}\left(  W\right)
,\delta^{\prime\prime}\left(  W\right)  ,\delta^{\prime\prime\prime}\left(
W\right)  ,$ are the \textquotedblleft kinematic equations\textquotedblright%
\ while those derived from the coefficients of $\delta\left(  W\right)  $ are
the \textquotedblleft dynamical equations\textquotedblright\ for each field.
These kinematic equations do not contain interactions (see however new
generalization in Eq.(\ref{homF})) and are in precise agreement\footnote{See
section IV-A in \cite{2tGravDetails} for a more detailed discussion of the
relation of these equations to Sp(2,R).} with Eqs.(\ref{sp3}-\ref{sp5})
derived from the Sp$\left(  2,R\right)  $ algebra of the $Q_{ij}\left(
X,P\right)  $.

The action (\ref{action}) has its own new gauge symmetries. It was shown in
\cite{2tGravity}\cite{2tGravDetails} that by fixing some of these gauge
symmetries, and solving the kinematic (not the dynamic!) equations for all the
fields, $W,\Omega,G^{MN},$ the physics of the 2T action may be reduced to 1T
physics in various shadows. The conformal shadow, which is the most familiar
to particle physicists, is then described by an effective action in $d$
dimensions which takes the conventional form of a conformally coupled scalar
$\phi\left(  x\right)  $ (shadow of $\Omega\left(  X\right)  $) interacting
with the gravitational field $g_{\mu\nu}\left(  x\right)  $ (shadow of
$G_{MN}\left(  X\right)  $) as follows
\begin{equation}
S\left(  g,\phi\right)  =\int d^{d}x\sqrt{-g}\left(  \frac{1}{2}g^{\mu\nu
}\partial_{\mu}\phi\partial_{\nu}\phi+a_{d}R\phi^{2}-V\left(  \phi\right)
\right)  . \label{confScalar}%
\end{equation}
For the special value of $a_{d}$ given above, it is well known that this
action has a local scaling Weyl symmetry $S(\tilde{g},\tilde{\phi})=S\left(
g,\phi\right)  $ under the gauge transformation
\begin{equation}
\tilde{g}_{\mu\nu}\left(  x\right)  =e^{2\lambda\left(  x\right)  }g_{\mu\nu
}\left(  x\right)  ,\;\tilde{\phi}\left(  x\right)  =e^{-\frac{d-2}{2}%
\lambda\left(  x\right)  }\phi\left(  x\right)  . \label{Weyl}%
\end{equation}
Tracing back the origin of this symmetry, it was discovered that it emerges as
a remnant of the general coordinate transformations in the extra dimensions,
while not being a Weyl symmetry of the original action \cite{2tGravDetails}.
Hence this is an accidental symmetry of the conformal shadow. With this
symmetry, the role of the \textit{shadow} $\phi\left(  x\right)  $ is that of
a \textquotedblleft compensator\textquotedblright\ as known in conventional 1T
conformal symmetry\footnote{I emphasize that the \textit{shadow} $\phi\left(
x\right)  $ may be fixed to a constant, but the original field $\Omega\left(
X\right)  $ is not a constant, it still depends on the extra dimensions beyond
$x^{\mu}$ in a specific way.}. It can be gauge fixed to a constant
$\phi\left(  x\right)  \rightarrow\phi_{0}$ if so desired, thus obtaining the
1T-physics of pure Einstein gravity with an undetermined cosmological
constant
\begin{equation}
S\left(  g,\phi_{0}\right)  =\frac{1}{2\kappa_{d}^{2}}\int d^{d}x\sqrt
{-g}\left[  R\left(  g\right)  -\Lambda_{d}\right]  , \label{shadowGrav}%
\end{equation}
where the Newton and cosmological constants in $d$ dimensions are determined
by $\phi_{0},\lambda$
\begin{equation}
\frac{1}{2\kappa_{d}^{2}}=a_{d}\phi_{0}^{2},\;\;\frac{\Lambda_{d}}{2\kappa
_{d}^{2}}=V\left(  \phi_{0}\right)  =\lambda\frac{d-2}{2d}\phi_{0}^{\frac
{2d}{d-2}}.
\end{equation}

Now we come to one of the main points of the present paper concerning the
interactions of matter with the gravitational field in the 2T formulation. The
Sp(2,R)-consistent rules for matter fields of Klein-Gordon, Dirac and
Yang-Mills types in $d+2$ dimensions were given in \cite{2tGravity}. For
scalar fields the consistent rules turn out to be more general as compared to
\cite{2tGravity} a follows.

The Sp$\left(  2,R\right)  $-consistent action is unique when there is only
one scalar field $\Omega,$ as in (\ref{action}). When there are more scalar
fields $\Phi_{m}\left(  X\right)  ,$ $m=1,2,3\cdots$ (all arranged into real
fields), I will show in the next section that the interactions include a
function $U\left(  \Phi\right)  ,$ a potential energy $V\left(  \Phi\right)  $
and a sigma-model type metric $g_{mn}\left(  \Phi\right)  ,$ with some
constraints on them, as follows%
\begin{equation}
S=\gamma\int d^{d+2}X~\sqrt{G}\left\{
\begin{array}
[c]{l}%
\delta\left(  W\right)  [a_{d}U\left(  \Phi\right)  R\left(  G\right)
-\frac{1}{2}g_{mn}\left(  \Phi\right)  G^{MN}\partial_{M}\Phi^{m}\partial
_{N}\Phi^{n}-V\left(  \Phi\right)  ]\\
+\delta^{\prime}\left(  W\right)  [a_{d}U\left(  \Phi\right)  \left(
4-\nabla^{2}W\right)  +a_{d}\partial W\cdot\partial U\left(  \Phi\right)  ]
\end{array}
\right\}  ~ \label{actionU}%
\end{equation}
where $a_{d}=\frac{d-2}{8\left(  d-1\right)  }.$ For a single field, $U,V,g$
are unique as given in Eq.(\ref{action}), $U=\Omega^{2},$ $g_{\Omega\Omega
}=-1,$ and $V=\lambda\frac{d-2}{2d}\Omega^{\frac{2d}{d-2}}.$

Some general physics requirement are as follows. $\left(  i\right)  $ First,
the metric in field space, $g_{mn}\left(  \Phi\right)  ,$ must not have any
zero eigenvalues for generic values of the fields $\Phi_{m};$ this insures
that all scalar fields have a kinetic term. $\left(  ii\right)  $ Second, each
positive eigenvalue of $g_{mn}$ corresponds to a physical scalar field while
each negative eigenvalue corresponds to a negative norm ghost. At least one
negative eigenvalue occurs as seen in the example of the single field
$U=\Omega^{2}.$ For each additional negative eigenvalue the full theory must
have some extra gauge symmetries beyond the accidental Weyl symmetry of
Eq.(\ref{Weyl}) to remove negative norm scalars. $\left(  iii\right)  $ Third,
$U\left(  \Phi\right)  $ should satisfy certain positivity conditions that
insure that gravity is an attractive (not repulsive) force, at least in the
patch of spacetime that makes up our universe, as will be discussed below.

To illustrate these points it is useful to review the example of $U,g_{mn}$
that was given in \cite{2tGravity}
\begin{align}
U\left(  \Phi\right)   &  =\Omega^{2}-\sum_{i=1}^{N}S_{i}^{2}\;\;\text{ ,
\ \ }\label{exU}\\
g_{mn}\left(  \Phi\right)   &  =-\frac{1}{2}\frac{\partial^{2}U}{\partial
\Phi^{m}\partial\Phi^{n}}=diag\left(  -1,+1,\cdots,1\right)  , \label{exg}%
\end{align}
where $\Phi_{0}\equiv\Omega,\;\Phi_{i}\equiv S_{i}.$ In this example the
metric $g_{mn}$ and the function $U\left(  \Phi\right)  $ have SO$\left(
N,1\right)  $ symmetry. This symmetry as well as the form of $U\left(
\Phi\right)  $ emerges uniquely in 2T supergravity if one requires that each
scalar has the standard canonical normalization of its kinetic term (i.e.
constant metric $g_{mn}$). To understand the physical effects of the negative
eigenvalue consider the conformal shadow analogous to Eq.(\ref{confScalar}) in
which all shadow scalars are conformally coupled to gravity with the special
value of $a_{d}$ and the quadratic $U\left(  \Phi\right)  $ of Eq.(\ref{exU}).
This special coupling of scalars to $R$ insures that in the conformal shadow
there is an overall local Weyl symmetry that rescales all \textit{shadow
}scalars equally (as in \ref{Weyl}). This Weyl symmetry is essential both to
generate the Newton constant and to remove one negative norm ghost as
explained in \cite{2tGravity}\cite{2tGravDetails}\cite{2tcosmology}. Indeed,
the field $\Omega$ has an extra minus sign in the kinetic term as compared to
the fields $S_{i}.$ This sign of $\Omega$ is the sign required in the
conformal shadow of Eq.(\ref{confScalar}) in order to obtain a positive Newton
constant upon the gauge fixing of the Weyl symmetry, as in
Eq.(\ref{shadowGrav}). However, this sign is the wrong sign for the kinetic
energy, and it makes the shadow $\phi$ of $\Omega$ a negative norm ghost
field. This is no problem since the Weyl gauge symmetry removes this ghost
from the spectrum when the gauge is fixed to obtain the Einstein action of
Eq.(\ref{shadowGrav}). The additional scalars $S_{i}\left(  X\right)  $ must
have the positive eigenvalue of $g_{mn}$ in the kinetic term so that their
shadows $s_{i}\left(  x\right)  $ are physical, positive norm, scalars. For
this reason, in the expression of $U\left(  \Phi\right)  =\Omega^{2}-\sum
S_{i}^{2}$ there must be a relative minus sign between $\Omega$ and the other
scalars $S_{i}.$ Now, one notices that the full $U\left(  \Phi\right)  $
rather than only $\Omega^{2}$ plays the role of an effective Newton constant
$16\pi G=\left(  a_{d}U\left(  \Phi\right)  \right)  ^{-1}=\left(  a_{d}%
\Omega^{2}-a_{d}\sum S_{i}^{2}\right)  ^{-1}$. Hence one must consider some
positivity requirements that limits the fields $\Phi_{m}\left(  X\right)  $ to
the region $U\left(  \Phi\right)  >0$ in field space in order to have a
positive $G$ that results in an attractive force for gravity.

What could go wrong if the dynamics of the scalar fields, including various
choices for the potential energy\footnote{The most general $V\left(
\Phi\right)  $ compatible with (\ref{exU},\ref{exg}) is homogeneous and may be
written as $V\left(  \Phi\right)  =\Omega^{\frac{2d}{d-2}}v\left(  \frac
{S_{i}}{\Omega}\right)  ,$ where $v\left(  x_{i}\right)  $ is any function of
its arguments \cite{2tGravity}\cite{2tGravDetails}. \label{hom2}} $V\left(
\Phi\right)  ,$ permit field configurations in which the effective Newton
constant $16\pi G=\left(  a_{d}U\left(  \Phi\right)  \right)  ^{-1}$ switches
sign? A conservative approach to prevent physical disasters is to demand
positivity; in particular in the quadratic example in Eq.(\ref{exU}) one may
require a field space that satisfies $\Omega^{2}>\sum_{i=1}^{N}S_{i}^{2}.$
However, this is an artificial condition that may be violated by the dynamics
of the coupled field equations. More interesting is to investigate the physics
of what happens if the dynamics lets $U\left(  \Phi\right)  $ evolve to the
configuration $U\left(  \Phi\right)  =0$ and even switch sign. In regions of
spacetime where $U\left(  \Phi\right)  $ is negative there would be
effectively antigravity (repulsive forces) rather than gravity (attractive forces).

In the conformal shadow familiar to particle physicists antigravity has not
been observed, so what are the observable effects if the positivity condition
is not obeyed by the dynamics? This question was investigated in a simplified
and exactly solvable cosmological model \cite{2tcosmology}, where it was shown
that $U\left(  \Phi\right)  =0$ corresponds to the Big Bang, while the region
$U\left(  \Phi\right)  <0$ is a pre-Big-Bang region that is not accessible in
our own universe, thus avoiding phenomenological problems. Interesting
cosmological questions arise for realistic and complete models, such as
whether our spacetime region of the universe is compatible with the existence
of various other antigravity regions of spacetime in the early or later epochs
during the evolution of the universe? Answering such questions have a bearing
in a fundamental theory on which forms of $U\left(  \Phi\right)  $ are
consistent with the physics we observe.

Having explained the nature of the physical issues that arise through the
quadratic example $U\left(  \Phi\right)  =[\Omega^{2}-\sum S_{i}^{2}],$ I will
next prove that the form of the action in Eq.(\ref{actionU}) is required
generally for consistency with the Sp$\left(  2,R\right)  $ homothety
constraints (\ref{sp3}-\ref{sp5}) on the geometry $G_{MN}\left(  X\right)  $
in $d+2$ dimensions. Afterwards I will discuss more general forms of $U\left(
\Phi\right)  $ and the additional gauge symmetries required when the
corresponding metric $g_{mn}\left(  \Phi\right)  $ in field space has more
than one negative eigenvalue.

At this point it is important to note that the emergent shadow field theory
that follows from (\ref{actionU}) is a special one among all the possible 1T
field theories containing scalar fields coupled to gravity. For example, when
$U\left(  \Phi\right)  =[\Omega^{2}-\sum S_{i}^{2}],$ the emergent shadow
\cite{2tGravity}\cite{2tGravDetails} similar to (\ref{shadowGrav}) including
the shadows of the $\Omega,S_{i}$, requires that all scalar fields (such as
Higgs, grand unified generalizations, scalar superpartners in SUSY theories,
inflaton, etc.) must couple to $R$ quadratically with the same coefficient
$a_{d}$ up to $\pm$ signs (see e.g. \cite{2tGravity}\cite{2tGravDetails}).
This is allowed but not motivated by a principle in a generic 1T field theory.
In this paper we learned that there is some freedom in the choice of $U\left(
\Phi\right)  ,g_{mn}\left(  \Phi\right)  $, but this freedom is further
reduced when additional requirements, such as symmetries in supergravity, are
considered, as will be explained in section (\ref{more}). After expressing the
conformal shadow (familiar to particle physicists) in the Einstein
frame\footnote{The Einstein frame is obtained by a Weyl transformation in the
conformal shadow (recall this is really a general coordinate transformation in
the extra dimensions \cite{2tGravDetails})$.$ See e.g. \cite{2tcosmology} for
the quadratic example of Eq.(\ref{exU}), in which this Weyl gauge amounts to
eliminating the dilaton $\phi$ in favor of the physical fields $s_{i}$, such
as $\phi=\pm\left[  1/2\kappa_{d}^{2}+\sum s_{i}^{2}\right]  ,$ where $\left(
2\kappa_{d}^{2}\right)  ^{-1}$ is the Newton constant and $\phi,s_{i}$ are the
conformal shadows of the original fields $\Omega,S_{i}.$} the tight relations
among interactions produced by the functions $U\left(  \Phi\right)
,g_{mn}\left(  \Phi\right)  $ could be physically distinguishable from a
comparable generic 1T field theory that may not motivate the same type of
restrictions. At the present time \textit{elementary} scalar fields in
particle physics (such as the Higgs particle) have not been constrained by
experiment. So an immediate test of certain patterns of scalar couplings in 1T
field theory, as motivated by 2T field theory, is not available, but this
could change in the future.

\section{2T-gravity action and consistency with Sp$\left(  2,R\right)  $}

To justify that the proposed action (\ref{actionU}) is the most general form
for 2T field theory of interacting scalars and gravity, I now compare the
Sp$\left(  2,R\right)  $ equations in Eq.(\ref{sp3}-\ref{sp5}) to the
\textit{kinematic} equations that result from the variation of the action. The
variation $\delta S$ has the form of Eq.(\ref{vary}) except that instead of
the single scalar field $\Omega$ there are now many fields $\Phi^{m}.$ The
kinematic equations are those proportional to $\delta^{\prime}\left(
W\right)  ,\delta^{\prime\prime}\left(  W\right)  ,\delta^{\prime\prime\prime
}\left(  W\right)  $, so I concentrate on those terms symbolized in
Eq.(\ref{vary}) by the letters $\left(  B_{MN}^{G},C_{MN}^{G};B^{\Phi_{m}%
},C^{\Phi_{m}};A^{W},B^{W},C^{W}\right)  $, all of which must vanish to
extremize the action.

Using similar steps to Refs.\cite{2tGravity}\cite{2tGravDetails} I derive
these coefficients. First note that the coefficients $C_{MN}^{G},C^{\Phi_{m}%
},C^{W}$ all have the same common factor $\left(  G^{MN}\partial_{M}%
W\partial_{N}W-4W\right)  .$ The only way that all $C$-coefficients vanish is
for the field $W\left(  X\right)  $ to satisfy the equation
\begin{equation}
G^{MN}\partial_{M}W\partial_{N}W=4W. \label{sp1}%
\end{equation}
This matches precisely the first two equations in Eq.(\ref{sp3}) that follow
from the Sp$\left(  2,R\right)  $ worldline gauge symmetry, once the vector
$V^{M}$ is identified as
\begin{equation}
V^{M}=\frac{1}{2}G^{MN}\partial_{N}W. \label{V}%
\end{equation}

To arrive at this result it is important to emphasize that the same $U\left(
\Phi\right)  $ must appear in all three terms in the action (\ref{actionU})
where they are indicated. One could have started with three different
functions $U_{1}\left(  \Phi\right)  ,U_{2}\left(  \Phi\right)  ,U_{3}\left(
\Phi\right)  ,$ in those three terms and then find out that they must be the
same $U\left(  \Phi\right)  $ because otherwise there would be additional
terms to cancel, proportional to $\partial_{M}W\partial_{N}W$ or $\partial
_{M}W\partial_{N}U_{i}$ in $B_{MN}^{G}$ and $C_{MN}^{G},$ that would be
inconsistent with the Sp$\left(  2,R\right)  $ constraints (\ref{sp3}%
-\ref{sp5}). So, $U_{1}=U_{2}=U_{3}=U$ is another consequence of demanding
consistency with the Sp$\left(  2,R\right)  $ constraints (\ref{sp3}-\ref{sp5}).

Next examine the coefficients $B_{MN}^{G},B^{\Phi_{m}},B^{W},A^{W}$ that are
given by \cite{2tGravity}
\begin{align}
0  &  =B_{MN}^{G}=a_{d}U\left(  \Phi\right)  \left[  G_{MN}\left(
-6+\nabla^{2}W+\partial W\cdot\partial\ln U\left(  \Phi\right)  \right)
-\nabla_{M}\partial_{N}W\right]  ,\label{BG}\\
0  &  =B^{\Phi_{m}}=g_{mn}\left(  \Phi\right)  \partial W\cdot\partial\Phi
^{n}+2a_{d}\frac{\partial U}{\partial\Phi^{m}}\left(  6-\nabla^{2}W\right)
,\label{BF}\\
0  &  =B^{W}=a_{d}\left[  U\left(  \Phi\right)  \left(  16-2\nabla
^{2}W\right)  -2\partial W\cdot\partial U\left(  \Phi\right)  \right]
,\label{BW}\\
0  &  =A^{W}=a_{d}U\left(  \Phi\right)  R\left(  G\right)  -2a_{d}\nabla
^{2}U\left(  \Phi\right)  -\frac{1}{2}g_{mn}\left(  \Phi\right)  \partial
\Phi^{m}\cdot\partial\Phi^{n}-V\left(  \Phi\right)  . \label{AW}%
\end{align}
where $\cdot$ means contraction with $G^{MN}$ and $\nabla^{2}$ is the
Laplacian constructed with the metric $G^{MN}.$ By contracting Eq.(\ref{BG})
with $G^{MN}$ and using $G^{MN}G_{MN}=d+2,$ one can derive an equation for
$\nabla^{2}W$%
\begin{equation}
\left(  d+2\right)  \left(  -6+\partial W\cdot\partial\ln U\left(
\Phi\right)  \right)  +\left(  d+1\right)  \nabla^{2}W=0.
\end{equation}
Combining this equation with the $B^{W}=0$ equation (\ref{BW}), the two
unknown quantities $\nabla^{2}W$ and $\partial W\cdot\partial\ln U\left(
\Phi\right)  $ are uniquely determined as
\begin{equation}
\nabla^{2}W=2\left(  d+2\right)  ,\;\partial W\cdot\partial\ln U\left(
\Phi\right)  =-2\left(  d-2\right)  . \label{kinWU}%
\end{equation}
Plugging this result back into Eq.(\ref{BG}) one finds%
\begin{equation}
G_{MN}=\nabla_{M}V_{N}=\frac{1}{2}\nabla_{M}\partial_{N}W=\frac{1}{2}\left(
\partial_{M}\partial_{N}W+\Gamma_{MN}^{P}\partial_{P}W\right)  . \label{sp2}%
\end{equation}
This is precisely equivalent to the homothety condition on the geometry
required by the Sp$\left(  2,R\right)  $ constraints (\ref{sp3}-\ref{sp5}).
There remains dealing with the kinematic equations $B^{\Phi_{m}}=A^{W}=0$ of
Eqs.(\ref{BF},\ref{AW}).

With the results (\ref{sp1},\ref{sp2}) that follow from the action, so far it
is evident that the 2T field theory action principle proposed in
Eq.(\ref{actionU}) is the most general one compatible with the underlying
Sp$\left(  2,R\right)  $ constraints (\ref{sp3}-\ref{sp5}) on the geometry, as
well as general coordinate invariance, without higher order derivatives of the
spacetime metric $G_{MN}$. This unique action has also led to the unique
kinematic conditions on $W\left(  X\right)  $ and $U\left(  \Phi\right)  $
shown in Eq.(\ref{kinWU}). The later condition on $U\left(  \Phi\right)  $ may
be written as a homothety condition $V^{M}\partial_{M}U\left(  \Phi\left(
X\right)  \right)  =-\left(  d-2\right)  U\left(  \Phi\left(  X\right)
\right)  ,$ where the left side is the Lie derivative for a scalar field using
the vector $V^{M}$ in Eq.(\ref{V}).

Next solve Eq.(\ref{BF}), $B^{\Phi_{m}}=0,$ after substituting $\left(
6-\nabla^{2}W\right)  =-2\left(  d-1\right)  $ as follows%
\begin{equation}
\partial W\cdot\partial\Phi^{n}=\frac{1}{2}\left(  d-2\right)  g^{nm}\left(
\Phi\right)  \frac{\partial U}{\partial\Phi^{m}}. \label{homF}%
\end{equation}
This kinematic constraint is also a generalized homothety condition on the
fields $\Phi^{n}$ which follow from Sp$\left(  2,R\right)  $ in the presence
of interactions provided by $g^{nm}\frac{\partial U}{\partial\Phi^{m}}$. In
the simple quadratic example of Eqs.(\ref{exU},\ref{exg}), the homothety
condition (\ref{homF}) for the scalars $\Phi^{m}$ becomes simply,
$V^{M}\partial_{M}\Phi^{m}=-\frac{1}{2}\left(  d-2\right)  \Phi^{m},$ which is
the Sp$\left(  2,R\right)  $ kinematic condition familiar from previous
studies of 2T-physics either in the BRST approach \cite{2tbrst2006} or the
flat spacetime 2T field theory approach \cite{2tstandardM}. In flat spacetime
, with $V^{M}=X^{M}$ (see footnote \ref{kinEqs}) this equation simply means
that $\Phi\left(  X\right)  $ is homogeneous $\Phi\left(  tX\right)
=t^{-\frac{1}{2}\left(  d-2\right)  }\Phi\left(  X\right)  .$ So, for more
general $U\left(  \Phi\right)  ,g_{mn}\left(  \Phi\right)  ,$ the kinematic
equation (\ref{homF}) should be understood as the generalized homothety or
\textquotedblleft homogeneity\textquotedblright\ condition, including
interactions. Solving this equation in a convenient choice of spacetime
coordinates in curved space (see examples in \cite{2tGravity}%
\cite{2tGravDetails}) reduces the field dependence on spacetime $X^{M}$ by one
coordinate. Since Eq.(\ref{homF}) is a first order differential equation in a
single coordinate, it can always be solved exactly for any interaction
$g^{nm}\frac{\partial U}{\partial\Phi^{m}}$.

Next, multiply both sides of Eq.(\ref{homF}) by $\frac{\partial\ln U}%
{\partial\Phi^{n}}.$ After summing over $n,$ the left hand side becomes
$\partial W\cdot\partial\ln U\left(  \Phi\right)  ;$ and using its derived
value in Eq.(\ref{kinWU}), the right hand side of (\ref{homF}) yields
\begin{equation}
g^{nm}\left(  \Phi\right)  \frac{\partial U}{\partial\Phi^{m}}\frac{\partial
U}{\partial\Phi^{n}}=-4U\left(  \Phi\right)  . \label{g1}%
\end{equation}
It is interesting to note the similarity of this equation to Eq.(\ref{sp1}),
although one is in field space $\Phi^{m}$ while the other is in position space
$X^{M}$. In a similar way one can also obtain the following equation from
(\ref{BF}) by multiplying both sides of (\ref{homF}) with $\partial
W\cdot\partial\Phi^{m}$%
\begin{equation}
g_{mn}\left(  \Phi\right)  \partial W\cdot\partial\Phi^{m}\partial
W\cdot\partial\Phi^{n}=-\left(  d-2\right)  ^{2}U\left(  \Phi\right)  .
\label{g11}%
\end{equation}
It should be noted that Eqs.(\ref{g1},\ref{g11}) are regarded as conditions on
the metric $g_{mn}\left(  \Phi\right)  $ in field space, which restrict the
types of possible interactions of the scalars for a given $U\left(
\Phi\right)  .$

An example of a metric $g_{mn}\left(  \Phi\right)  $ and a $U\left(
\Phi\right)  $ that satisfy all of these constraints Eqs.(\ref{homF}%
,\ref{g1},\ref{g11}) is the quadratic example given in Eqs.(\ref{exU}%
,\ref{exg}). In this example the metric is constant and both the kinetic term
and the $U\left(  \Phi\right)  R$ terms in the action have SO$\left(
N,1\right)  $ symmetry. Furthermore, the homothety condition \ref{homF} takes
the simple form $\partial W\cdot\partial\Phi^{m}=-\left(  d-2\right)  \Phi
^{m},$ which is the curved space generalization of the Sp$\left(  2,R\right)
$ constraint given in footnote \ref{kinEqs}, and is easily solved
\cite{2tGravity}\cite{2tGravDetails}.

Another example of $g_{mn}\left(  \Phi\right)  $ and $U\left(  \Phi\right)  $
that satisfy the constraints in Eqs.(\ref{homF},\ref{g1},\ref{g11}) is%
\begin{equation}
U\left(  \Phi\right)  =\Omega^{2},\;\;g^{mn}=\left(
\begin{array}
[c]{cc}%
-1 & -\frac{S^{j}}{\Omega}\\
-\frac{S^{i}}{\Omega} & \left(  g^{ij}-\frac{S^{i}S^{j}}{\Omega^{2}}\right)
\end{array}
\right)  ,\;g_{mn}=\left(
\begin{array}
[c]{cc}%
\left(  -1+\frac{S^{k}g_{kl}S^{l}}{\Omega^{2}}\right)  & -\frac{g_{jk}S^{k}%
}{\Omega}\\
-\frac{g_{il}S^{l}}{\Omega} & g_{ij}%
\end{array}
\right)  , \label{compensator}%
\end{equation}
where the $N\times N$ sub-metric $g_{ij}(\Omega,S)$ is an arbitrary metric in
field space. In this second example $U\left(  \Phi\right)  =\Omega^{2}$ is
positive definite, while the Sp$\left(  2,R\right)  $ homothety constraint
(\ref{homF}) takes the form $\partial W\cdot\partial\Phi^{m}=-\left(
d-2\right)  \Phi^{m}$ for $\Phi^{m}=\left(  \Omega,S^{i}\right)  ,$ which is
the same as the other example.

Finally, there is the equation $A^{W}=0$ in (\ref{AW}). Actually, this is not
a kinematic equation, but rather it is a dynamical equation since second order
spacetime derivatives appear. To analyze this equation we need to take into
account the dynamical equations $A_{MN}^{G}=0$ and $A^{\Phi_{m}}=0$ for all
the fields $\Phi^{m}$ and $G_{MN}.$ It turns out that $A^{W}=0$ is
automatically satisfied provided the dynamical equations $A_{MN}^{G}=0$ and
$A^{\Phi_{m}}=0$ are satisfied (see \cite{2tGravity}\cite{2tGravDetails}), so
this is not an additional constraint to contend with.

\section{More symmetry and constraints on scalars \label{more}}

More constraints on the scalar couplings $U\left(  \Phi\right)  ,V\left(
\Phi\right)  ,g_{mn}\left(  \Phi\right)  $ can arise because of stronger
symmetries. Specific examples of this occurs with 2T supersymmetry
\cite{susy2tN1}\cite{susy2tN2N4} which restricts the form of $V\left(
\Phi\right)  ,$ and 2T supergravity \cite{sugra1}\cite{sugra2} which restricts
the form of $g_{mn}\left(  \Phi\right)  $ to be a function constructed from
$U\left(  \Phi\right)  .$ Without going into the details of the 2T
supersymmetry, it is possible to understand the effects of supersymmetry on
the scalars in 2T supergravity by considering the conformal shadows of the
scalars that are expected to appear in conventional 1T SUSY and 1T
supergravity theories.

For example, for conventional $\mathcal{N}$=1 SUSY in 4-dimensions, the main
effect on $V\left(  \Phi\right)  $ is that it must be constructed from complex
fields (chiral multiplets) in the form of D-terms and F-terms with a
holomorphic superpotential $f\left(  \Phi\right)  $, in a well known form that
we don't need to elaborate on here (for reviews see \cite{sugra3BaggerWess}%
\cite{sugra4Weinberg}).

More constraints are found in supergravity. For example, for conventional
$\mathcal{N}$=1 supergravity in 4-dimensions, there is a K\"{a}hler potential
coupled to $R$ that also determines the metric in field space that occurs in
the kinetic term of complex chiral multiplets (see e.g. formulas 31.6.57 to
31.6.61 in \cite{sugra4Weinberg}).

From such 1T shadows of 2T supergravity, with various numbers of supercharges
$\mathcal{N},$ it is staightforward to deduce the corresponding constraints on
$g_{mn}\left(  \Phi\right)  ,U\left(  \Phi\right)  $ in 2T supersymmetric
field theory, beyond the constraints already described in the previous
section. We will not be specific here for various $\mathcal{N}$, but only
indicate that one typical constraint is that $g_{mn}\left(  \Phi\right)  $ is
constructed from $U\left(  \Phi\right)  $ as a second derivative in field
space, such as%
\begin{equation}
g_{mn}\left(  \Phi\right)  =-\frac{1}{2}\frac{\partial^{2}U\left(
\Phi\right)  }{\partial\Phi^{m}\partial\Phi^{n}}. \label{realMetric}%
\end{equation}
Actually, the constraint in $\mathcal{N}$=1 supergravity is even stronger in
terms of complex fields that yield a K\"{a}hler metric in field space
\begin{equation}
g_{m\bar{n}}\left(  \Phi,\bar{\Phi}\right)  =-\frac{\partial^{2}U\left(
\Phi,\bar{\Phi}\right)  }{\partial\Phi^{m}\partial\bar{\Phi}^{\bar{n}}}.
\end{equation}
Rather than the specific form (complex or real), the derivative form is what
we wish to pursue here to make the following observations. For this reason we
will stick to real fields and the metric in (\ref{realMetric}) to maintain a
consistent notation with the previous sections (the result will be similar for
complex fields).

The derivative form of the metric permits a different approach to solving the
homothety constraint (\ref{BF}) on the scalars. Inserting Eq.(\ref{realMetric}%
) into (\ref{BF}) and using $\left(  6-\nabla^{2}W\right)  =-2\left(
d-1\right)  $ as before, $B^{\Phi^{m}}=0$ takes the form
\begin{equation}
-\frac{1}{2}\frac{\partial^{2}U\left(  \Phi\right)  }{\partial\Phi^{m}%
\partial\Phi^{n}}\partial W\cdot\partial\Phi^{n}-\frac{1}{2}\left(
d-2\right)  \frac{\partial U}{\partial\Phi^{m}}=0.
\end{equation}
Rather than solving this in the form of (\ref{homF}), which still holds, the
chain rule leads to a simpler result, namely%
\begin{equation}
\partial W\cdot\partial\frac{\partial U}{\partial\Phi^{m}}+\left(  d-2\right)
\frac{\partial U}{\partial\Phi^{m}}=0.
\end{equation}
This homothety constraint is a linear equation in $\frac{\partial U}%
{\partial\Phi^{m}}$ and, other than being in curved space, it looks the same
as the Sp$\left(  2,R\right)  $ kinematic constraint for the scalar field in
footnote (\ref{kinEqs}). Combined with the result in (\ref{kinWU}), $\partial
W\cdot\partial\ln U\left(  \Phi\right)  =-2\left(  d-2\right)  ,$ this implies
that
\begin{equation}
\partial W\cdot\partial\Phi^{m}=-\left(  d-2\right)  \Phi^{m},
\end{equation}
and that $U\left(  \Phi\right)  $ and $V\left(  \Phi\right)  $ must be
homogeneous of degree $2$ and $2d/\left(  d-2\right)  $ respectively in field
space%
\begin{equation}
U\left(  t\Phi\right)  =t^{2}U\left(  \Phi\right)  ,\;V\left(  t\Phi\right)
=t^{\frac{2d}{d-2}}V\left(  \Phi\right)  . \label{hom22}%
\end{equation}
Hence, when the metric is constructed as a second derivative of $U\left(
\Phi\right)  ,$ as in 2T supergravity, an additional consequence is that
$U\left(  \Phi\right)  $ and $V\left(  \Phi\right)  $ are homogeneous as
indicated. Furthermore $U\left(  \Phi\right)  $ must also satisfy the
non-linear equation that follows from Eq.(\ref{g1})
\begin{equation}
\frac{\partial U}{\partial\Phi^{m}}\left(  \frac{\partial\Phi\otimes
\partial\Phi}{\partial^{2}U}\right)  ^{mn}\frac{\partial U}{\partial\Phi^{n}%
}=2U\left(  \Phi\right)  . \label{sugraCond}%
\end{equation}

These are a lot of constraints on $U\left(  \Phi\right)  ,$ as well as
$g_{mn}\left(  \Phi\right)  ,$ that eliminate previously possible solutions.
Nevertheless there still remains some freedom. Note that a homogeneous
$U\left(  \Phi\right)  $ of degree $2$ does not necessarily mean quadratic,
since ratios of fields may also occur in $U\left(  \Phi\right)  $. Some
additional symmetry conditions, such as global or local gauge symmetries that
must be respected in the full theory can narrow down the possibilities. For
example, asking for a global SO$\left(  N,1\right)  $ symmetry in the kinetic
term completely nails down both $U\left(  \Phi\right)  $ and $g_{mn}\left(
\Phi\right)  $ to have the quadratic form in Eq.(\ref{exU},\ref{exg}).
Alternatively, asking for no particular symmetry of $U\left(  \Phi\right)  $
but only asking for a canonical normalization of the kinetic term with a
constant metric $g_{mn}$ also nails down $U\left(  \Phi\right)  $ to be the
same quadratic of Eq.(\ref{exU}).

As an example, applying this quadratic case for complex scalars $\Phi^{m}%
,\bar{\Phi}^{m},$ to construct a 2T $\mathcal{N}$=1 supergravity theory in
$4+2$ dimensions \cite{2tSugra} yields the following potential energy
$V\left(  \Phi,\bar{\Phi}\right)  $ when $U$ is given by $U\left(  \Phi
,\bar{\Phi}\right)  =\left(  \Phi^{0}\bar{\Phi}^{0}-\sum\Phi^{i}\bar{\Phi}%
^{i}\right)  $
\begin{equation}
V\left(  \Phi,\bar{\Phi}\right)  =\frac{\partial f\left(  \Phi\right)
}{\partial\Phi^{m}}\frac{\partial\bar{f}\left(  \bar{\Phi}\right)  }%
{\partial\bar{\Phi}^{n}}g^{mn},\;f\left(  t\Phi\right)  =t^{3}f\left(
\Phi\right)  . \label{Vnoscale1}%
\end{equation}
Here $g^{mn}$ is the constant SU$\left(  N,1\right)  $ metric\footnote{When
both the kinetic terms and $U$ are all rewritten in terms of real fields, the
symmetry of both $U$ and the kinetic terms is actually SO$\left(  2N,2\right)
,$ with two negative eigenvalues of the metric when rewritten in a real basis.
But, because of the complex nature of the superpotential, what appears in $V$
is the metric for the subgroup SU$\left(  N,1\right)  .$ \label{SO(N,2)}} in
Eq.(\ref{exg}), and $f\left(  \Phi\right)  $ is the analytic superpotential
which must be homogeneous of degree 3. Again this does not necessarily mean
that $f\left(  \Phi\right)  $ is cubic, since ratios of fields may also occur
(in non-renormalizable effective theories). Although this looks like a simple
$F$-term in the potential, I emphasize that this is the full form after all
the supergavity machinery, including the K\"{a}hler potential is taken into
account as explained in \cite{2tSugra}. The simplicity occurs because of the
special form of the quadratic function $U\left(  \Phi,\bar{\Phi}\right)  .$
Note that due to the indefinite metric $g_{mn}$, this $V$ is not a positive
definite potential energy. Having a negative contribution to the potential
energy, despite the exact local supersymmetry, is typical in supergravity. An
additional positive definite D-term is added to this potential energy in the
standard form when Yang-Mills gauge fields are coupled to supergravity (see
e.g. \cite{sugra3BaggerWess}\cite{sugra4Weinberg}).

I have shown that in 2T field theory there are a variety of constraints on
scalar fields. The first and foremost are the Sp$\left(  2,R\right)  $
constraints on the overall theory which results in kinematic equations on all
the fields of all spins. These kinematic equations are related to the gauge
symmetries that remove ghosts and make the theory unitary directly in $d+2$
dimensions. Their effect on scalar fields is twofold. First, the scalar field
must obey a kinematic equation, whose most general form, including
interactions, is given in Eq.(\ref{homF}). The solutions of the kinematic
equations correspond to the shadows in 1T field theory and help interpret the
predictions of 2T physics in the language of 1T physics. Second, the possible
interactions of scalars among themselves $V\left(  \Phi\right)  $ and with
gravity described by $U\left(  \Phi\right)  ,g_{mn}\left(  \Phi\right)  $ must
obey certain conditions, including especially (\ref{g1},\ref{g11}) which are
regarded as restrictions on the metric in field space. Supersymmetry puts
further severe constraints on $V\left(  \Phi\right)  ,$ and $g_{mn}$
,$U\left(  \Phi\right)  $ in the form (\ref{hom22},\ref{sugraCond}), while
gauge symmetries and global symmetries of the overall theory narrow down the
possible interactions.

The patterns of scalar field interactions predicted by 2T physics for the 1T
conformal shadow (familiar setting in particle physics) can also be introduced
in 1T field theory by hand, but are not necessarily motivated by a similar principle.

\section{Epilogue: No scale models and the cosmological constant}

In a separate paper I will discuss 2T supergravity. The present study of
scalars was motivated by trying to solve some puzzles for how to
supersymmetrize 2T gravity. The essential issue was that the accidental Weyl
symmetry (\ref{Weyl}) in the conformal shadow was crucial to remove the ghost
(the negative eigenvalue in the metric $g_{mn}\left(  \Phi\right)  $)
associated with the field $\Omega$. Recall that the conformal shadow of
$\Omega$ played the role of a conformal compensator familiar in 1T field
theory. To supersymmetrize, this ghost compensator could be made a member of a
chiral multiplet, and hence there would be an additional ghost due to the
complex field nature of the compensator chiral multiplet. Not only that, there
would also be fermionic partners of these ghosts. I was puzzled for a long
time how these additional bosonic and fermionic ghosts could be removed
consistently with supersymmetry, and was wondering if supersymmetrizing 2T
gravity would require a different approach than chiral
multiplets?\footnote{For example, the linear multiplet \cite{linMultiplet}
does not require the complexification, but after all it is equivalent to the
chiral multiplet.}

I thank S. Ferrara for clarifying some aspects of gauge symmetries in
supergravity \cite{sugraConf}\cite{sugraConf2}. Just one brief remark on
SU$\left(  2,2|1\right)  $ was sufficient to show me the way and solve all the
puzzles as follows. In 2T SUSY field theory with $\mathcal{N}$ supersymmetries
in 4+2 dimensions there is a global SU$\left(  2,2|\mathcal{N}\right)  $
symmetry \cite{susy2tN1}\cite{susy2tN2N4}. The SU$\left(  2,2\right)  $ part
of it is the linearly realized SO$\left(  4,2\right)  $ of the 4+2 dimensions.
To construct $\mathcal{N}$=1 supergravity, just as SO$\left(  4,2\right)  $ is
turned into a local symmetry, the full SU$\left(  2,2|1\right)  $ must also
become local. If such a 2T supergravity theory exists in 4+2 dimensions it
must be that the conformal shadow in 3+1 dimensions is also locally symmetric
under SU$\left(  2,2|1\right)  .$ Indeed, Ferrara pointed out that such a
formulation of 1T Poincar\'{e} supergravity was considered some time ago in a
conformal formalism \cite{sugraConf}\cite{sugraConf2}. The SU$\left(
2,2|1\right)  $ has all the local symmetries to remove all the ghosts
associated with the compensator. Specifically, the local superconformal
S-supersymmetry removes the fermionic member of the compensator chiral
multiplet, the gauged U$\left(  1\right)  $ R-symmetry (with a non-propagating
auxiliary vector field) removes the phase of the remaining complex boson, and
finally the local Weyl symmetry fixes the compensator to the Newton constant
as in Eq.(\ref{shadowGrav}). From SU$\left(  2,2|1\right)  $ there remains
only the local Lorentz and the local Q-supersymmetry, which are the evident
local symmetries of Poincar\'{e} supergravity! Hence the road to 2T
supergravity is clear.

The details of the 2T supergravity will be given elsewhere \cite{2tSugra}, but
here I outline the resulting conformal shadow with particular emphasis on the
scalars. The shadow of the $\mathcal{N}$=1 Poincar\'{e} supergravity in 3+1
dimensions contains the following supermultiplets: the graviton supermultiplet
$\left(  e_{\mu}^{a},\psi_{\mu}^{\alpha},b_{\mu},z\right)  $ where $b_{\mu
},z(\equiv s+ip)$ are auxiliary fields \cite{sugraConf}, plus the chiral
supermultiplets $\left(  \phi^{m},\psi^{m},F^{m}\right)  $ labelled by
$m=0,1,\cdots,N,$ where $F^{m}$ are auxiliary fields. The bosonic part of the
Lagrangian that can be compared to conventional supergravity is\footnote{See
e.g. Eq.(31.6.57) in \cite{sugra4Weinberg}, where the Newton constant terms
are dropped, and the auxiliary fields $b_{\mu},s,p$ are renormalized by a
convenient numerical factor of 2/3. Note also that the factor of $1/6$ in
front of $R$ comes from $2a_{d}=\frac{1}{6}$ for $d=4$ (coming from $d+2=6$).
Here we have $2a_{d}$ instead of $a_{d}$ because of the complex basis for the
fields. Similarly, the factor of $3$ in front of $3\bar{z}f\left(
\phi\right)  $ more generally is given as $\frac{d+2}{2}.$}%
\[
\frac{1}{e}\mathcal{L}_{bose}=\left\{
\begin{array}
[c]{c}%
U\left(  \phi,\bar{\phi}\right)  \left(  \frac{1}{6}R\left(  g\right)
-z\bar{z}+g^{\mu\nu}b_{\mu}b_{\nu}\right) \\
+\frac{\partial^{2}U}{\partial\phi^{m}\partial\bar{\phi}^{n}}\left(  g^{\mu
\nu}\partial_{\mu}\phi^{m}\partial_{\nu}\bar{\phi}^{n}-F^{m}\bar{F}^{n}\right)
\\
+\left[  -\left(  zF^{m}+ib^{\mu}\partial_{\mu}\phi^{m}\right)  \frac{\partial
U}{\partial\phi^{m}}+\frac{\partial f}{\partial\phi^{m}}F^{m}+3\bar{z}f\left(
\phi\right)  \right]  +c.c.
\end{array}
\right\}
\]
where $f\left(  \phi\right)  $ is the superpotential. The multiplet labelled
by $m=0$ contains the field $\phi^{0}\left(  x\right)  $ that plays the role
of the \textit{complex} \textquotedblleft compensator\textquotedblright\ so it
describes a supermultiplet of negative norm ghosts. There is just the required
amount of gauge symmetry to remove them from the physical spectrum, and
generate from them the Newton constant, as described above. In particular the
U$\left(  1\right)  $ gauge field that removes the phase of $\phi^{0}\left(
x\right)  $ is the auxiliary field $b_{\mu}.$ To exhibit the U$\left(
1\right)  $ gauge symmetry associated with $b_{\mu}$ I define the following
covariant derivative with a non-linear action of the $U\left(  1\right)  $
transformation (it becomes linear oly for quadratic $U$)
\begin{equation}
D_{\mu}\phi^{m}=\partial_{\mu}\phi^{m}+ib_{\mu}\frac{\partial U}{\partial
\bar{\phi}^{n}}\left(  \frac{\partial\bar{\phi}\otimes\partial\phi}%
{\partial^{2}U}\right)  ^{nm},
\end{equation}
where the last factor is the inverse of the K\"{a}hler metric. Then this
Lagrangian takes the U$\left(  1\right)  $ gauge invariant form%
\[
\frac{1}{e}\mathcal{L}_{bose}=\left\{
\begin{array}
[c]{c}%
\frac{1}{6}U\left(  \phi,\bar{\phi}\right)  R\left(  g\right)  +\frac
{\partial^{2}U}{\partial\phi^{m}\partial\bar{\phi}^{n}}g^{\mu\nu}D_{\mu}%
\phi^{m}D_{\nu}\bar{\phi}^{n}+\left(  \frac{\partial\bar{\phi}\otimes
\partial\phi}{\partial^{2}U}\right)  ^{nm}\frac{\partial\bar{f}}{\partial
\bar{\phi}^{n}}\frac{\partial f}{\partial\phi^{m}}\\
+\frac{\partial^{2}U}{\partial\phi^{m}\partial\bar{\phi}^{n}}\left(  F+\bar
{z}\phi+\left(  \frac{\partial\phi\otimes\partial\bar{\phi}}{\partial^{2}%
U}\right)  \frac{\partial\bar{f}}{\partial\bar{\phi}}\right)  ^{m}\left(
\bar{F}+z\bar{\phi}+\left(  \frac{\partial\bar{\phi}\otimes\partial\phi
}{\partial^{2}U}\right)  \frac{\partial f}{\partial\phi}\right)  ^{n}\\
-\bar{z}\left(  \phi^{m}\frac{\partial f}{\partial\phi^{m}}-3f\left(
\phi\right)  \right)  -z\left(  \bar{\phi}^{m}\frac{\partial\bar{f}}%
{\partial\bar{\phi}^{m}}-3f\left(  \bar{\phi}\right)  \right)
\end{array}
\right\}  .
\]
This shows that gauge invariance is possible only when $U\left(  \phi
,\bar{\phi}\right)  $ satisfies the non-linear condition in
Eq.(\ref{sugraCond}). The gauge invariance is required in order to remove the
negative norm ghost.

The last two lines in this form vanish after integrating out the auxiliary
fields $F^{m}$ and $z$. In any case the last line does not even appear because
2T supegravity already requires in Eq.(\ref{Vnoscale1}) that $f\left(
\phi\right)  $ had to be homogeneous of degree $3;$ hence the auxiliary field
$z$ dropped out automatically anyway. Hence the bosonic Lagrangian simplifies
greatly to the form\footnote{This form is invariant under the following
infinitesimal Weyl transformation that generalizes Eq.(\ref{Weyl}):
$\delta_{\lambda}g_{\mu\nu}=2\lambda\left(  x\right)  g_{\mu\nu}$ and
$\delta_{\lambda}\phi^{m}=-\frac{d-2}{2}\lambda\left(  x\right)  g^{mn}%
\frac{\partial U}{\partial\bar{\phi}^{n}},$ and $\delta_{\lambda}b_{\mu}=0,$
where $g^{mn}\equiv\left(  \frac{\partial\phi\otimes\partial\bar{\phi}%
}{\partial^{2}U}\right)  ^{mn}.$ Note that the Weyl transformation of
$\phi^{m}$ is derived from the Sp$\left(  2,R\right)  $ condition on the
parent field $\Phi^{m}$ in $d+2$ dimensions in Eq.(\ref{homF}). As explained
in \cite{2tGravDetails}, this implies that the Weyl trasformation for all the
fields in the shadow amounts to a reparametrization of the coordinates in the
extra dimensions. There is no Weyl symmetry in the higher dimensional theory.
Note also that this Weyl symmetry holds even for the more general case of
$g_{mn}\left(  \phi,\bar{\phi}\right)  $ that satisfies the complex version of
Eq.(\ref{sugraCond}), $\frac{\partial U}{\partial\phi^{m}}g^{mn}\frac{\partial
U}{\partial\bar{\phi}^{n}}=-U,$ even when $g_{mn}$ is not constructed from
derivatives of $U$ (i.e. not taking into account the local supersymmetry
conditions in supergravity). For any such $g_{mn}\left(  \phi,\bar{\phi
}\right)  ,$ after integrating out the gauge field $b_{\mu},$ the kinetic term
for the scalars takes the following form $g_{mn}D\phi^{m}\cdot D\bar{\phi}%
^{n}=g^{\mu\nu}\left(  g_{mn}\partial_{\mu}\phi^{m}\partial_{\nu}\bar{\phi
}^{n}-J_{\mu}J_{\nu}\right)  $, where $J_{\mu}=\frac{1}{2}\left(
i\partial_{\mu}\phi^{m}\frac{\partial\ln U}{\partial\phi^{m}}-i\partial_{\mu
}\bar{\phi}^{m}\frac{\partial\ln U}{\partial\bar{\phi}^{m}}\right)  .$}
\begin{equation}
\frac{1}{e}\mathcal{L}_{bose}=\frac{1}{6}U\left(  \phi,\bar{\phi}\right)
R\left(  g\right)  +\frac{\partial^{2}U}{\partial\phi^{m}\partial\bar{\phi
}^{n}}g^{\mu\nu}D_{\mu}\phi^{m}D_{\nu}\bar{\phi}^{n}+\left(  \frac
{\partial\bar{\phi}\otimes\partial\phi}{\partial^{2}U}\right)  ^{nm}%
\frac{\partial\bar{f}}{\partial\bar{\phi}^{n}}\frac{\partial f}{\partial
\phi^{m}}.\label{shadowSugra}%
\end{equation}
The potential energy is then
\begin{equation}
V\left(  \phi,\bar{\phi}\right)  =-\left(  \frac{\partial\bar{\phi}%
\otimes\partial\phi}{\partial^{2}U}\right)  ^{nm}\frac{\partial\bar{f}%
}{\partial\bar{\phi}^{n}}\frac{\partial f}{\partial\phi^{m}}.
\end{equation}
as given in Eq.(\ref{Vnoscale1}). Note that this is homogeneous of degree 4,
$V\left(  t\phi,t\bar{\phi}\right)  =t^{4}U\left(  \phi,\bar{\phi}\right)  $.
Recall that here we also require that $U\left(  \phi,\bar{\phi}\right)  $ is
homogeneous of degree $2,$ and that it must satisfy the complex version of the
non-linear condition (\ref{sugraCond})%
\begin{equation}
U\left(  t\phi,t\bar{\phi}\right)  =t^{2}U\left(  \phi,\bar{\phi}\right)
,\;\;\frac{\partial U}{\partial\phi^{m}}\left(  \frac{\partial\phi
\otimes\partial\bar{\phi}}{\partial^{2}U}\right)  ^{mn}\frac{\partial
U}{\partial\bar{\phi}^{n}}=U\left(  \phi,\bar{\phi}\right)  .\label{Ueq}%
\end{equation}
Only the $U\left(  \phi,\bar{\phi}\right)  $ that can solve these equations
(with non-zero eigenvalues in the K\"{a}hler metric $g_{mn}=-\frac
{\partial^{2}U}{\partial\phi^{m}\partial\bar{\phi}^{n}}$) are admitted in the
3+1 dimensional conformal shadow of 2T supergravity in 4+2 dimensions.

A $U\left(  \phi,\bar{\phi}\right)  $ that satisfies Eq.(\ref{Ueq}), together
with an analytic homogeneous superpotential that satisfies $f\left(
t\phi\right)  =t^{3}f\left(  \phi\right)  $, determine fully the scalar field
interactions in 2T supergravity and its conformal shadow given in
Eq.(\ref{shadowSugra}). \textit{This is the restriction on scalars in
1T-physics that arises from 2T-supergravity.}

An example of a $U\left(  \phi,\bar{\phi}\right)  $ that satisfies these
equations is the complex version of the quadratic case in Eq.(\ref{exU}) that
I discussed several times in this paper. In a complex basis it has the form
\begin{equation}
\text{example: }U\left(  \phi,\bar{\phi}\right)  =\phi^{0}\bar{\phi}^{0}%
-\sum_{i=1}^{N}\phi^{i}\bar{\phi}^{i}=-g_{mn}\phi^{m}\bar{\phi}^{n}.
\label{quad}%
\end{equation}
This constant metric $g_{mn}=diag\left(  -1,+1,\cdots,+1\right)  $ leads to
canonicaly normalized complex scalars, with an automatic linearly realized
SU$\left(  N,1\right)  $ global symmetry in the kinetic and $R$ terms of the
action (\ref{shadowSugra}). This symmetry may be broken by the choice of the
superpotential $f\left(  \phi\right)  $. In this example, the action
(\ref{shadowSugra}) requires that all scalars must be conformally coupled to
gravity. Note that not only the compensator, but all scalars are conformally
coupled. This is possible in 1T field theory, but it is not motivated by a
principle, like it is in 2T-physics.

It is worth mentioning that the work on 2T supergravity has led in a natural
way to a class of no scale models of 1T supergravity that are a good starting
point for understanding the basic problem of the smallness of the cosmological
constant \cite{noScale1}. In particular the quadratic $U$ of Eq.(\ref{quad})
immediately produces an attractive no scale model with the potential energy
$V$ given above in Eq.(\ref{Vnoscale1}), as detailed in \cite{2tSugra}. It is
a coincidence that right after resolving the 2T supergravity puzzles, and
having constructed the quadratic model, a brief discussion with C. Kounnas
whom I ran into unexpectedly has attracted my attention to the no scale ideas
\cite{noScale1}-\cite{quevedo} for which the 2T supergravity path is quite
natural. Specifically, a no scale model is obtained from the above 2T
supergravity approach simply by taking the following basis for the fields
$\phi^{\pm}=\left(  \phi^{0}\pm\phi^{N}\right)  /\sqrt{2},$ and then writing
the potential (\ref{Vnoscale1}) in the form
\begin{align}
V\left(  \phi,\bar{\phi}\right)   &  =-\frac{\partial f}{\partial\phi^{0}%
}\frac{\partial\bar{f}}{\partial\bar{\phi}^{0}}+\frac{\partial f}{\partial
\phi^{N}}\frac{\partial\bar{f}}{\partial\bar{\phi}^{N}}+\sum_{i=1}^{N-1}%
\frac{\partial f}{\partial\phi^{i}}\frac{\partial\bar{f}}{\partial\bar{\phi
}^{i}},\nonumber\\
&  =-\frac{\partial f}{\partial\phi^{+}}\frac{\partial\bar{f}}{\partial
\bar{\phi}^{-}}-\frac{\partial f}{\partial\phi^{-}}\frac{\partial\bar{f}%
}{\partial\bar{\phi}^{+}}+\sum_{i=1}^{N-1}\frac{\partial f}{\partial\phi^{i}%
}\frac{\partial\bar{f}}{\partial\bar{\phi}^{i}}. \label{Vnoscale2}%
\end{align}
If one takes a superpotential $f\left(  \phi\right)  $ that depends only
$\phi^{+}$ and $\phi^{i},$ i.e. $\frac{\partial f}{\partial\phi^{-}}=0,$ then
in the remaining $V\left(  \phi,\bar{\phi}\right)  $ each term is strictly
positive for all values of the fields. For the scalars, the minimum of the
potential $V\left(  \phi,\bar{\phi}\right)  $ can only occur when each term
vanishes at field values that satisfies $\frac{\partial f}{\partial\phi^{i}%
}=0$ for $i=1,\cdots,N-1,$ while $\frac{\partial f}{\partial\phi^{+}}%
$=anything (since $\frac{\partial f}{\partial\phi^{-}}=0$). Therefore the
absolute minimum of the potential is necessarily at zero $V_{\min}=0$,
yielding automatically a vanishing cosmological constant even after
spontaneous breakdown of symmetries that cause phase transitions in the
history of the universe (such as electroweak, SUSY, grand unification,
inflation, etc.)\footnote{This discussion is before one goes to the Einstein
frame. The Einstein frame is most easily reached by making convenient Weyl and
U$\left(  1\right)  $ gauge choices for the conformal shadow of $\Phi^{+},$
namely, $\operatorname{Im}\left(  \phi^{+}\right)  =0$ and $\operatorname{Re}%
\phi^{+}=\left(  \phi^{-}+\bar{\phi}^{-}\right)  ^{-1}\left(  \frac{1}%
{2\kappa^{2}}+\sum\phi_{i}\bar{\phi}_{i}\right)  $ for the quadratic example.
After a similar gauge choice for any $U,$ and using the homogeneity of the
potential $V,$ the effective potential in the Einstein frame is written as
$V\left(  \phi,\bar{\phi}\right)  =\left(  \operatorname{Re}\phi_{+}\right)
^{4}V\left(  \frac{\phi}{\phi^{+}},\frac{\bar{\phi}}{\bar{\phi}^{+}}\right)
=V_{eff}\left(  \phi^{+},\phi_{i},\bar{\phi}^{+},\bar{\phi}_{i}\right)  .$
Hence they are proportional to each other $V_{eff}\sim V$ with an overall
positive coefficient $\left(  \operatorname{Re}\phi^{+}\right)  ^{4}$.
Therefore the discussion of the minumum of the potential is the same in the
Einstein frame. Furthermore, including the D-terms do not change this
discussion because D-terms are strictly positive and they must vanish
separately at the minimum of the potential. \label{fixf+}}.

The field $\phi^{+}$ can be gauge fixed conveniently as in footnote
(\ref{fixf+}) by using the Weyl gauge symmetry in Eq.(\ref{Weyl}), thus
generating the Newton constant $\left(  2\kappa^{2}\right)  ^{-1}$. In this
specific gauge the quadratic example can be compared to the no scale model
discussed in \cite{noScale2}. The homogeneous superpotential of
Eq.(\ref{Vnoscale1}) may be written as $f\left(  \phi\right)  =\left(
\phi^{+}\right)  ^{3}\rho\left(  \phi^{i}/\phi^{+}\right)  ,$ with an
arbitrary $\rho\left(  z^{i}\right)  .$ This $\rho\left(  z^{i}\right)  $ may
be chosen to fit particle physics phenomenology, including SUSY breaking,
which may be clarified in experiments at the LHC if the SUSY scale is within
its reach\footnote{An interesting modification of the quadratic form is
$U\left(  \phi,\bar{\phi}\right)  =\alpha\phi^{+}\bar{\phi}^{+}-\phi^{+}%
\bar{\phi}^{-}-\phi^{-}\bar{\phi}^{+}+\sum\phi_{i}\bar{\phi}_{i}$ with any
constant $\alpha.$ This gives a non-diagonal but constant $g_{mn}$ and a
potential $V\left(  \phi,\bar{\phi}\right)  =-\frac{\partial f}{\partial
\phi^{+}}\frac{\partial\bar{f}}{\partial\bar{\phi}^{-}}-\frac{\partial
f}{\partial\phi^{-}}\frac{\partial\bar{f}}{\partial\bar{\phi}^{+}}-\alpha
\frac{\partial f}{\partial\phi^{-}}\frac{\partial\bar{f}}{\partial\bar{\phi
}^{-}}+\sum_{i=1}^{N-1}\frac{\partial f}{\partial\phi_{i}}\frac{\partial
\bar{f}}{\partial\bar{\phi}_{i}}.$ When $\frac{\partial f}{\partial\phi^{-}%
}=0$ this still reduces to a no scale model with the strictly positive $V$,
but a different $U$ for any $\alpha.$ The parameter $\alpha$ can be used for
phenomenological purposes as discussed elsewhere.}. The remaining field
$\phi^{-}$ is unfixed at the minimum of the potential at the classical level
(a flat direction, hence \textit{no scale}). Quantum corrections can stabilize
the remaining $\phi^{-}$ field. There exist schemes \cite{quevedo} that may
explain the smallness of the observed cosmological constant after the quantum corrections.

One lesson of the $\mathcal{N}$=1 supergravity example above is that
$g_{mn}\left(  \Phi\right)  $ can have more than one negative eigenvalues (see
footnote \ref{SO(N,2)}). To kill the corresponding additional ghosts, there
must be Yang-Mills type gauge symmetries, such as the U$\left(  1\right)  $
R-symmetry in the SU$\left(  2,2|1\right)  $, or its generalizations for
higher $\mathcal{N}$. Furthermore higher $\mathcal{N}$ supergravity admits
gauge symmetries with non-propagating auxiliary vector fields. Such gauge
symmetries, combined with the Weyl symmetry, are then used to remove all the
negative norm ghost scalar fields, leaving behind the familiar scalar fields
in 1T supergravity theories that are described as moduli in coset spaces of
certain non-compact U-duality groups. The benefit of keeping the negative norm
scalars in the initial formulation of the 2T supergravity theory is to make
evident hidden symmetries and then using the gauge symmetries in the most
convenient way (example \cite{2tcosmology}) to analyze the physics in the 1T shadows.

The 2T approach has been indicating in many settings, including gravity and
supergravity in this paper, that there is an ambient d+2 dimensional spacetime
in which the fundamental form of the theory resides. The shadows in d
dimensions distort the fundamental form of the equations, just like an
observer's choice of coordinates in general relativity inserts a distortion.
Unlike general relativity, in 2T-physics this distortion leads to different
choices of \textquotedblleft time\textquotedblright\ in 1T-physics, and hence
to different 1T-physics interpretations. The conventional formulation of 1T
physics is just one of the shadows, namely the conformal shadow familiar in
particle physics. This familiarity is the reason to concentrate mainly on the
conformal shadow in many of the discussions because this helps to digest the
physical meaning of 2T-physics at least in one familiar setting. However, the
benefits of the 2T formulation will be mainly in exploring the other shadows
and in using the duality relationships among the shadows to develop useful
computational techniques as well as new insights about the meaning of space
and time, as discussed partially in \cite{emergentfieldth1}%
\cite{emergentfieldth2}\cite{rivelles}. In this regard, Weinberg's recent
results for Green's functions \cite{weinberg}, which amount to Green's
functions in flat 4+2 dimensional 2T field theory including the standard model
\cite{2tstandardM}, is one of the explicit examples that can be explored by
using the reduction techniques to various shadows as suggested in footnote
(\ref{weinbergNote}).

On the fundamental theory side, 2T physics for strings and branes and 2T
superfield theory in higher dimensions is still underdeveloped (for their
status see \cite{phaseSpace}). I note that SUSY Yang-Mills theory has already
been constructed in 10+2 dimensions as will be presented in the near future
\cite{susyYM102}. Further exploration of the fundamentals in 2T physics along
these lines should lead next to supergravity in 10+2 and 11+2 dimensions, thus
providing a 2T version of M theory.

$\allowbreak$

\begin{acknowledgments}
I thank Sergio Ferrara, Costas Kounnas, J.P. Derendinger, Fernando Quevedo,
Kellog Stelle, Eugene Cremmer, Dieter L\"{u}st, Hans Nilles, Stuart Raby,
Emannuel Floratos and Juan Maldacena for very helpful and supportive
discussions on 1T and 2T supergravity and no scale models. I also thank CERN,
the ICTP, and the LPTENS XL\`{e}me Institut d'\'{E}t\'{e}, for providing
stimulating atmospheres where this research was conducted.
\end{acknowledgments}


\begin{thebibliography}{99}                                                                                               %


\bibitem {phaseSpace}I. Bars, \textquotedblleft Gauge Symmetry in Phase Space,
Consequences for Physics and Spacetime\textquotedblright, arXiv:1004.0688
[hep-th], Lecture in honor of Murray Gell-Mann, to appear in IJMPA.

\bibitem {rivelles}J.E. Frederico and V.O. Rivelles, \textquotedblleft The
Transition Amplitude for 2T Physics\textquotedblright, arXiv:1002.1263 [hep-th].

\bibitem {weinberg}S. Weinberg, \textquotedblleft Six-dimensional Methods for
Four-dimensional Conformal Field Theories\textquotedblright, arXiv:1006.3480
[hep-th]. See footnote \ref{weinbergNote} in the present paper for some comments.

\bibitem {waldron}R. Bonezzi, E. Latini, A. Waldron, \textquotedblleft
Gravity, Two Times, Tractors, Weyl Invariance and Six Dimensional Quantum
Mechanics\textquotedblright, arXiv:1007.1724 [hep-th].

\bibitem {2tbacgrounds}I. Bars, \textquotedblleft Two time physics with
gravitational and gauge field backgrounds", Phys. Rev. \textbf{D62}, 085015
(2000) [arXiv:hep-th/0002140]; I. Bars and C. Deliduman, \textquotedblleft%
\ High spin gauge fields and two time physics", Phys. Rev. \textbf{D64},
045004 (2001) [arXiv:hep-th/0103042].

\bibitem {2tGravity}I. Bars, \textquotedblleft Gravity in
2T-Physics\textquotedblright, Phys. Rev. \textbf{D77} (2008) 125027
[arXiv:0804.1585 [hep-th]].

\bibitem {2tGravDetails}I. Bars and S-H Chen, \textquotedblleft Geometry and
Symmetry Structures in 2T Gravity\textquotedblright, Phys. Rev. D79 (2009)
085021 [arXiv:0811.2510 (hep-th)].

\bibitem {spin2t}I. Bars and C. Deliduman, Phys. Rev. \textbf{D58} (1998)
106004, hep-th/9806085.

\bibitem {2Tfields2000}I. Bars, \textquotedblleft Two time physics in field
theory\textquotedblright, Phys. Rev. \textbf{D62}, 046007 (2000) [arXiv:hep-th/0003100].

\bibitem {super2t}I. Bars, C. Deliduman and D. Minic, \textquotedblleft
Supersymmetric Two-Time Physics\textit{\textquotedblright, }Phys. Rev.
\textbf{D59} (1999) 125004, [arXiv:hep-th/9812161]; \textquotedblleft Lifting
M-theory to Two-Time Physics\textit{\textquotedblright, }Phys. Lett.
\textbf{B457} (1999) 275, [arXiv:hep-th/9904063].

\bibitem {2ttwistor}I. Bars, \textquotedblleft\ 2T physics formulation of
superconformal dynamics relating to twistors and supertwistors,"
Phys.\ Lett.\ B \textbf{483}, 248 (2000) [arXiv:hep-th/0004090].
\textquotedblleft Twistors and 2T-physics,\textquotedblright\ AIP Conf. Proc.
\textbf{767} (2005) 3 , [arXiv:hep-th/0502065].

\bibitem {twistorBP}I. Bars and M. Picon, \textquotedblleft Single twistor
description of massless, massive, AdS, and other interacting
particles,\textquotedblright\ Phys. Rev. \textbf{D73} (2006) 064002
[arXiv:hep-th/0512091]; \textquotedblleft Twistor Transform in d Dimensions
and a Unifying Role for Twistors,\textquotedblright\ Phys. Rev. \textbf{D73}
(2006) 064033, [arXiv:hep-th/0512348].

\bibitem {twistorLect}I. Bars, Lectures on Twistors, USC-06/HEP-B1, [arXiv:hep-th/0601091].

\bibitem {2tsuperstring}I. Bars, \textquotedblleft Twistor superstring in
2T-physics,\textquotedblright\ Phys. Rev. \textbf{D70} (2004) 104022, [arXiv:hep-th/0407239].

\bibitem {spinBO}I. Bars and B. Orcal, \textquotedblleft Generalized Twistor
Transform and Dualities, with a New Description of Particles With Spin, Beyond
Free and Massless\textquotedblright,\ arXiv:0704.0296 [hep-th].

\bibitem {2tbrst2006}I. Bars and Y-C. Kuo, \textquotedblleft Interacting
two-time Physics Field Theory with a BRST gauge Invariant
Action\textquotedblright, hep-th/0605267.

\bibitem {noncommutative}I. Bars, \textquotedblleft u$_{\ast}\left(
1,1\right)  $ non-commutative gauge theory as the foundation of 2T-physics in
field theory\textquotedblright, Phys. Rev. D64 (2001) 126001 [hep-th/0106013].
I. Bars and S. Rey, \textquotedblleft Noncommutative Sp(2,R) gauge theories: A
Field theory approach to two time physics.\textquotedblright, Phys. Rev. D64
(2001) 046005 [hep-th/0104135].

\bibitem {2tstandardM}I. Bars, \textquotedblleft The standard model of
particles and forces in the framework of 2T-physics\textquotedblright, Phys.
Rev. \textbf{D74} (2006) 085019 [arXiv:hep-th/0606045]. For a summary see
\textquotedblleft The Standard Model as a 2T-physics theory\textquotedblright, arXiv:hep-th/0610187.

\bibitem {susy2tN1}I. Bars and Y.C. Kuo, \textquotedblleft Field theory in
2T-physics with $N=1$ supersymmetry\textquotedblright\ Phys. Rev. Lett.
\textbf{99} (2007) 41801 [arXiv:hep-th/0703002]; \textit{ibid}.
\textquotedblleft Supersymmetric field theory in 2T-physics,\textquotedblright%
\ Phys. Rev. \textbf{D76 }(2007) 105028,. [arXiv:hep-th/0703002].

\bibitem {susy2tN2N4}I. Bars and Y.C. Kuo, \textquotedblleft N=2,4
Supersymmetric Gauge Field Theory in 2T-physics\textquotedblright\ Phys. Rev.
\textbf{D79} (2009) 025001 [arXiv:0808.0537].

\bibitem {emergentfieldth1}I. Bars, S-H. Chen and G. Quelin, \textquotedblleft
Dual Field Theories in (d-1)+1 Emergent Spacetimes from a Unifying Field
Theory in d+2 Spacetime,\textquotedblright\ Phys. Rev. \textbf{D76} (2007)
065016 [arXiv:0705.2834 [hep-th]].

\bibitem {emergentfieldth2}I. Bars, and G. Quelin, \textquotedblleft Dualities
among 1T-Field Theories with Spin, Emerging from a Unifying 2T-Field
Theory\textquotedblright, Phys. Rev. \textbf{D77} (2008) 125019
[arXiv:0802.1947 [hep-th]].

\bibitem {Dirac}P.A.M Dirac, Ann. Math. \textbf{37} (1936) 429.

\bibitem {kastrup}H. A. Kastrup, Phys. Rev. \textbf{150} (1966) 1183.

\bibitem {salam}G. Mack and A. Salam, Ann. Phys. \textbf{53} (1969) 174.

\bibitem {adler}S. Adler, Phys. Rev. \textbf{D6} (1972) 3445; \textit{ibid}.
\textbf{D8} (1973) 2400.

\bibitem {ferrara}S. Ferrara, A. F. Grillo, and R. Gatto, Ann. Phys. (NY) 76
(1973) 161; S. Ferrara, Nucl. Phys. \textbf{B77} (1974) 73.

\bibitem {fronsdal}F. Bayen, M. Flato, C. Fronsdal and A. Haidari, Phys. Rev.
\textbf{D32} (1985) 2673.

\bibitem {siegel}W. Siegel, Int. J. Mod. Phys. \textbf{A3} (1988) 2713; Int.
Jour. Mod. Phys. \textbf{A4} (1989) 2015.

\bibitem {vasiliev}C. R. Preitschopf and M. A. Vasiliev, Nucl. Phys.
\textbf{B549} (1999) 450 [arXiv:hep-th/9812113].

\bibitem {vasiliev2}M. A. Vasiliev, JHEP \textbf{12} (2004) 046, [hep-th/0404124].

\bibitem {2tcosmology}I. Bars and S-H Chen, \textquotedblleft The Big Bang and
Inflation United by an Analytic Solution,\textquotedblright\ e-Print: arXiv:1004.0752.

\bibitem {2tSugra}I. Bars, \textquotedblleft2T supergravity in 4+2
dimensions\textquotedblright, in preparation.

\bibitem {sugra1}S. Deser and B. Zumino, Phys. Lett. \textbf{62B} (1976) 335.

\bibitem {sugra2}D.Z. Freedman, P. van Nieuwenhuizen and S. Ferrara, Phys.
Rev. \textbf{D13} (1976) 3214.

\bibitem {sugra3BaggerWess}J.\symbol{126}Wess and J.\symbol{126}Bagger,
\textquotedblleft Supersymmetry and supergravity,\textquotedblright\ \{%
$\backslash$%
it Princeton Univ. Press (1992). \}

\bibitem {sugra4Weinberg}S. Weinberg, \textquotedblleft The quantum theory of
fields, Vol.III, supersymmetry\textquotedblright, \{%
$\backslash$%
it Camridge Univ. Press (2000)\}.

\bibitem {linMultiplet}For a review of the linear multiplet see: J.P.
Derendinger and F. Quevedo, \textquotedblleft The linear multiplet and quantum
four-dimensional string effective actions\textquotedblright, Nucl. Phys.
\textbf{B428} (1994) 282 [hep--th/9402007], and references therein.

\bibitem {sugraConf}S. Ferrara and P. van Nieuwenhuizen, \textquotedblleft The
auxiliary fields of supergravity\textquotedblright, Phys. Lett. \textbf{74B}
(1978) 333; \textit{ibid}. \textquotedblleft Tensor calculus for
supergravity\textquotedblright, Phys. Lett. \textbf{76B} (1978) 404.

\bibitem {stelleWest}K.S. Stelle and P.C. West, \textquotedblleft Minimal
auxiliary fields for supergravity\textquotedblright, Phys. Lett. \textbf{74B}
(1978) 330.

\bibitem {sugraConf2}S. Ferrara, M. Kaku, P. Townsend and P. van
Nieuwenhuizen, \textquotedblleft Unified field theories with U(N) internal
symmetries: gauging the superconformal group\textquotedblright, Nucl. Phys.
\textbf{B129} (1977) 125.

\bibitem {noScale1}E. Cremmer, S. Ferrara, C. Kounnas and D. Nanopoulos,
\textquotedblleft Naturally vanishing cosmological constant in N=1
supergravity\textquotedblright, Phys. Lett. \textbf{133B} (1983) 61.

\bibitem {noScale2}J. Ellis, C. Kounnas, and D. Nanopoulos, \textquotedblleft
No scale supersymmetric GUTs\textquotedblright, Nucl. Phys. \textbf{B247}
(1984) 373; \textit{ibid} \textquotedblleft No Scale Supergravity Models With
A Planck Mass Gravitino,\textquotedblright\ Phys. Lett.\{%
$\backslash$%
bf B143\} (1984) 410.

\bibitem {noScale3}A. Lahanas and D. Nanopoulos , \textquotedblleft The Road
to No Scale Supergravity,\textquotedblright\ Phys. Rept. 145 (1987) 1.

\bibitem {noScale4Kugo}T. Kugo and S. Uehara, \textquotedblleft\ Conformal and
Poincar\'{e} tensor calculi in N=1 supergravity,\textquotedblright\ Nucl.
Phys. \textbf{B226} (1983) 49; \textit{ibid}. \textquotedblleft\ Improved
superconformal gauge conditions in the N=1 supergravity Yang-Mills matter
system\textquotedblright, Nucl. Phys. \textbf{B222} (1983) 125; \textit{ibid}.
\textquotedblleft On the new nonminimal nersions of N=1 Poincar\'{e}
supergravity,\textquotedblright\ Nucl. Phys. \textbf{B226} (1983) 93.

\bibitem {quevedo}M. Cicoli, C.P. Burgess and F. Quevedo, \textquotedblleft
Fibre inflation: observable gravity waves from IIB string
compactifications\textquotedblright, JCAP \textbf{03} (2009) 013
[arXiv:0808.0691 [hep-th]].

\bibitem {susyYM102}I. Bars and Y.C. Kuo, \textquotedblleft Super Yang-Mills
theory (SYM) in 10+2 dimensions, linking SYM$_{d=4}^{N=4}$ and M(atrix) theory
to 2T-physics\textquotedblright, in preparation.
\end{thebibliography}
\end{document}